# Exceptional magneto-electric coupling and spontaneous electric polarization in anti-ferromagnet $Co_4Nb_2O_9$

P. Srivastava, S. Chaudhary, J. Saha and S. Patnaik

*School of Physical Sciences, Jawaharlal Nehru University, New Delhi-110067, India*

**Abstract**

Synthesis and extensive structural, pyroelectric, magnetic, dielectric and magneto-electric characterizations are reported for polycrystalline $Co_4Nb_2O_9$ towards unraveling the multiferroic state especially in reference to the magnetic "spin flop" transition. Magnetic measurements confirm the $Co_4Nb_2O_9$ becomes antiferromagnetic (AFM) at around 28 K but no clear evidence for spin-flop effect was found. Associated with the magnetic phase transition, a sharp peak in pyroelectric current indicates the appearance of the strong magneto-electric coupling below Neel temperature ($T_N$) with a large coupling constant upto 17.8 $\mu C/m^2 T$. Using temperature oscillation technique, we establish $Co_4Nb_2O_9$ to be a genuine multiferroic with spontaneous electric polarization in the anti-ferromagnetic state.

I. Introduction

Multiferroic materials promise several technological innovations based on coexisting ordered states of electric, elastic and magnetic phases.[1-4] In particular, a significant number of recently discovered multiferroic materials relate to magnetic structure driven ferroelectiricty where electric polarization is induced by magnetic structure (e.g. $TbMnO_3$)[2] instead of ab-initio non-centro symmetric crystal structure (e.g, $BaTiO_3$).[5] Towards, practical applications it is imperative that these prototypes possess the ferroelectric properties with high saturation polarization along with strong magneto-electric coupling. In this regard, the corundum compound $Co_4Nb_2O_9$ has attracted significant attention because of its exceptional Magneto-electric coupling parameter (30 ps/m).[6,7] However, its spontaneous polarization magnitude has remained controversial as the ferroelectricity in this compound was reported only in the presence of magnetic field-poled samples.[6] What it implies is that spontaneous polarization is too low to be of consequence. Furthermore, the anomalously high magneto-dielectric effect across the magnetic transition was assigned to a magnetic spin-flop transition. Using, highly resolved pyroelectric current oscillation technique, in this paper we establish that $Co_4Nb_2O_9$ possesses genuine ferroelectricity below its antiferromagnetic transition along with extremely high magneto-electric coupling parameter.

The magneto-electric materials (such as $Cr_2O_3$, $MnTiO_3$)[8,9] exhibit strong dependence of dielectric constant with external magnetic field. Several of these compounds are industrial compounds and are investigated intensely for novel physical properties towards next-generation spintronics applications. The type-II multiferroics, in which polarization exists only in a magnetically ordered state and is due to a particular type of magnetic structure, are theoretically expected to exhibit strong ME effect as well.[10] However, in the past decade, although magnetic control of polarization [8,11-20] has been widely reported and a good number of type-II multiferroics have been studied, [12,21-23] multiferroic compounds with large ME coupling have been found to be a rare phenomena. With regard to collinear antiferromagnet $Co_4Nb_2O_9$, it is reported that the compound has a α-$Al_2O_3$-type trigonal crystal structure with Neel temperature ~28K.[6] In this paper we report an investigation of ferroelectric state in the absence of magnetic field by measurement of pyroelectric current under thermal cycling in polycrystalline $Co_4Nb_2O_9$. In contrast to the previous study,[6] our results suggest that spontaneous polarization exists below Neel temperature even in the absence of magnetic field. More interestingly, a detailed measurement of the ME effect unveils a larger ME coupling constant along with coexisting spin-flop phase at low magnetic fields.

## II. Experimental:

Polycrystalline sample of $Co_4Nb_2O_9$ was synthesized by standard solid state reaction route. Stoichiometric amount of pure $Co_3O_4$ (Alfa Aesar, 98.0%), and $Nb_2O_5$ (Alfa Aesar, 98.0%), were used. The mixture was well ground for several hours in agate mortar-pestle. After grinding, the mixture was sintered at 900°C for 10 hrs in air. After the first sintering, the mixture was again well ground for few hours and then pressed in form of pellets (Diameter = 5 mm, thickness= 0.5 mm). These pellets were heated again in air at 1100°C for another 6 hrs. Both heating and cooling rates were kept at a rate of 5°C/min. Phase purity was confirmed at room temperature powder X-ray diffraction (Rigaku Ultima III X-ray diffractometer with Cu-K$_\alpha$ radiation). Lattice parameters were obtained from Rietveld refinement of the X-ray data using RIETAN 2000 software. The DC magnetization measurements were done in VSM mode of a *Cryogenic* Physical Property Measurement System (PPMS). The dielectric measurements were done using an Agilent E4980A LCR meter. For pyroelectric and dielectric measurements, electrodes were prepared on the sample by painting silver paste on the planar surfaces. The pyroelectric measurement was performed using a Kiethley 6514 electrometer and polarization was derived from the pyroelectric current by integrating over time. The sample was poled each time from well above ordering temperature under various combinations of electric and magnetic field. The poling field was removed at lowest temperature and terminals were sorted for 15 minutes to avoid any role of electrostatic stray charges. The warming ramp rate was fixed at 5 K/min. For dielectric measurements, silver paste was used to make contacts onto disc-shaped ceramic samples of 5 mm diameter and 0.5 mm thickness.

## III. Results and Discussion

Room temperature powder XRD pattern of polycrystalline $Co_4Nb_2O_9$ sample in the range of $2\theta = 10°$ to $90°$ is shown in Fig.1(a). The refinement done by Fullprof technique has been used for phase detection of synthesized sample. It is confirmed that all the spectra fit the standard database for the α-$Al_2O_3$-type trigonal type structure with space group $P\bar{3}c1$. From the XRD data, the lattice parameters are estimated to be a = 5.167 Å, c = 14.123 Å with α=90.00° and, γ =120.00°. X-ray diffraction data confirm almost phase pure synthesis of polycrystalline $Co_4Nb_2O_9$. The atomic coordinates are also estimated from the room temperature powder XRD and are summarized in Table 1. $Co_4Nb_2O_9$ crystallizes in α-$Al_2O_3$-type structure (space group P$\bar{3}$C1), which is schematically illustrated in Fig. 1(b). It is seen that two crystallographic sites for the Co ions are non-equivalent. The unit cell consists of two formula units of $Co_4Nb_2O_9$. Such corundum type of structure of general formula $A_4B_2O_9$ (A = Co, Fe, Mn, and B =Nb, Ta), were first prepared and investigated by Bertaut et al.[24] Recently, the magnetic structure of $Co_4Nb_2O_9$ material has been studied by neutron diffraction studies [24,25] and it is inferred that below Neel temperature, Co spins order parallel to the c-axis direction and create chains along the lines $(\frac{1}{3}, \frac{2}{3}, z)$ and $(\frac{2}{3}, \frac{1}{3}, z)$ with anti-parallel inter-chain coupling.[24] Further, strong analogies between the spin configuration of $Co_4Nb_2O_9$ and $Cr_2O_3$, imply that there is possibility to observe linear ME effect in $Co_4Nb_2O_9$ in consonance with properties of $Cr_2O_3$.[8]

The temperature dependence of magnetic susceptibility for $Co_4Nb_2O_9$ under various external magnetic fields is shown in Fig. 2(a). In the presence of external field, the susceptibility first increases with decreasing temperature. After reaching a maximum value, the susceptibility begins to drop upon further cooling, showing an antiferromagnetic behavior. The peak temperature around 28 K can be defined as the Neel temperature ($T_N$) of $Co_4Nb_2O_9$, which is consistent with the earlier report.[6,25] At all external magnetic fields, the susceptibility shows a distinctive behavior: it increases with further decrease in temperature and a bump around $T_N$ is observed, which is assigned to spin-flop transition of antiferromagnet. [6,26,27] According to Kolodiazhnyi, sufficiently large magnetic field applied away from the easy axis can lead to the spin-flop transition in $Co_4Nb_2O_9$, whereas magnetic field along the easy axis would stabilize the canted spin states. [25]The minimum required field to induce the spin flop transition in $Co_4Nb_2O_9$ is reported to be ~1.2 Tesla.[25] But we find unambiguous evidence for upturn in low temperature susceptibility (below Neel temperature) at sufficiently low field of about 200 G (shown in Fig. 2(a))[6,25]. Temperature dependence of inverse magnetic susceptibility, $\chi^{-1}$ (T), measured at H=0.1 T is shown in Fig. 2 (b). Linear fit of the $\chi^{-1}$ (T) dependence yields a Weiss temperature $\theta_W$ ≈ -51.68 K and Curie constant C=13.32 emu.K/mol.Oe (inset of Fig.2 (b)). We note that the spin flop feature is supposed to be most pronounced when the field is applied along c-axis of the crystal with the cobalt moments confined to basal plane of trigonal structure. In fact no spin flop feature is expected along 100 or 010 direction.[7] In our polycrystalline sample therefore, appearance of spin flop feature at fields 0.02 Tesla therefore puts question mark on the driving cause of high ME coupling in $Co_4Nb_2O_9$. Fig. 2 (c) plots the variation of magnetization as a function of external magnetic field for $Co_4Nb_2O_9$ (T = 4 K), which indicates a significant dominance of canted moments in the antiferromagnetic state. In fact, no upturn in magnetization is observed at field ~1.2 Tesla. In general, a large anisotropy leads to a significant spin-canting by increasing the possible anti-symmetric superexchange interaction, while a small anisotropy supports the spin-flop transition. [28,29] The magnetically induced polarization in conjunction with spin-flop phase has been reported in linear ME materials like $Cr_2O_3$, $MnTiO_3$, etc. [8,9] It is therefore interesting to investigate the electric polarization in $Co_4Nb_2O_9$ compound in relation to its magnetic structure.

Next we concentrate on the central question as to whether the induction of electric polarization necessarily requires the presence of magnetic field while cooling or not. In essence, there are several magneto-electric compounds which are not mutiferroic because of lack of spontaneous electric polarization. In order to confirm the existence of magnetism induced electric polarization in $Co_4Nb_2O_9$, we carry out the measurement of pyroelectric current, which is collected under various magnetic fields. Before the measurement, the sample was cooled from 35 to 8 K with an electric field of 200V/mm with a magnetic field of 0, 2, 4 and 5 T applied on the sample. Fig. 3(a) shows the measured pyroelectric current as a function of temperature and poled in magnetic field. In zero magnetic field, no clear signal of pyroelectric current is observed. However, when a magnetic field is applied, a pyroelectric current develops in a temperature range below the anti-ferromagnetic phase transition and the peak value increases with increasing magnetic fields. The temperature dependence of electric polarization, which is obtained by integration of pyroelectric current with respect to time is shown in Fig. 3 (b). An electric polarization of 89μC/$m^2$ is observed at 8 K with a poling. The inset of Fig. 3 (b) shows the magnetic field dependence of polarization at 8, 22 and 25K, respectively. It can be seen that the polarization increases with increasing magnetic field and above 2 Tesla, a linear ME effect is evidenced. [6,7,25] The ME coefficient $\alpha_{ME}$ = [P(H) - P(0)]/ΔH reaches up to 17.8 μC/$m^2$T at 5 T, indicating a large ME coupling. [6] Fig. 3 (c) shows the change in sign of polarization at all applied magnetic fields as a function of temperature with various positive and negative poling electric fields. A rather symmetric temperature dependence of polarization curve is observed, indicating that the electric polarization can be reversed by applying negative field.

In Fig. 4(a) we plot the pyroelectric current data following a temperature oscillation deep inside the antiferromagnetic state. We observe that the direction of pyroelectric current changes during cooling and heating cycles.[30,31] Clearly, when the temperature is varied from 25 K to 27K in a periodic manner, the changes in pyroelectric current follows the profile of temperature variation. We note that no magnetic field is applied during this measurement nor were it used for poling the sample. From Fig. 4 (a), we also observe that the pyroelectric current does not decay over time, in the absence of external magnetic field, which confirms the movement of permanent dipole. This is the evidence for the robust preservation of polarized state below the magnetic transition temperature. When heat is supplied to polarized dipoles, their arrangement tends towards the low ordered state and this is recovered when specimen temperature is lowered. Such reversible phenomenon leads to creation of opposite current in intrinsic ferroelectric material. Thus, in contrast to several other reports where no polarization was observed below $T_N$ in the absence of magnetic field.[6] We observe evidence for spontaneous polarization in the corundum $Co_4Nb_2O_9$ even in the absence of magnetic field. In Fig 4(b), we summarize the results of magnetic field dependent dielectric constant across the AFM transition in $Co_4Nb_2O_9$. For different applied magnetic fields, the temperature dependent dielectric constant ($\varepsilon$) in the vicinity of the AFM phase transition is shown. In the absence of magnetic field (H=0 T), dielectric constant of $Co_4Nb_2O_9$ is showing temperature independent behavior. In the presence of magnetic field, two essential features are observed, (i) a clear peak appears at the transition temperature and (ii) the height of the dielectric peak at the transition temperature increases with the magnetic fields. By applying magnetic field, the spin symmetry is broken and the peak becomes stronger with the increment of magnetic field. Magnetic field induced contribution to the dielectric constant, the magneto-dielectric effect $\Delta\varepsilon/\varepsilon(H=0)$ reaches up to 1 % at 5 T. Hence, the observation of the anomaly of $\varepsilon$ around the AFM ordering transition at $T_N$ = 28 K reconfirms strong magneto-dielectric feature of $Co_4Nb_2O_9$ material.

## IV. Conclusion

In summary, we have investigated the intrinsic multiferroic behavior in polycrystalline $Co_4Nb_2O_9$ sample with possible spin-flop phase. The experimental results reveal that spontaneous polarization arises below $T_N$ in the absence of magnetic field. A large magneto-electric coupling parameter (17.8 µC/m$^2$T at 5 T ) along with significant magneto-dielectric effects are observed although no evidence for spin-flop transition could be confirmed. The electric polarization increases with magnetic fields and is reversible by varying the polarity of the poling electric field. The microscopic origin for this rare phenomenon of coexisting strong magneto-electric coupling along with magnetic structure driven spontaneous electric polarization needs to be ascertained.

**Acknowledgements**

P. Srivastava acknowledges UGC, New Delhi-India for Dr. D. S. Kothari Post Doctoral Fellowship. We thank AIRF- JNU for magnetization measurements.

**Figure Captions:**

**Table 1.** Atomic coordinates derived from crystal structure analysis of $Co_4Nb_2O_9$.

**Fig.1 (a)** Room temperature powder XRD pattern of $Co_4Nb_2O_9$ polycrystalline sample with Reitveld refinement. The observed data (red), calculated line (black) and difference (blue) between the two are shown along with the Bragg positions (green). α-$Al_2O_3$-type trigonal structure with space group $P\bar{3}c1$ is confirmed. **(b)** Schematic crystal structure of $Co_4Nb_2O_9$.

**Fig.2 (a)** Temperature dependence of magnetic susceptibilities (arbitrary scale) under 0.02, 0.05, 0.1, and 1 T, external magnetic field for $Co_4Nb_2O_9$. **(b)** Temperature dependence of magnetic susceptibility at 0,1T with inset showing inverse susceptibility. Linear fit of the $\chi^{-1}(T)$ dependence yields Weiss temperature $\theta_W \approx$ - 51.68 K and Curie constant C=13.32 emu K/mol. Oe. **(c)** The magnetization versus magnetic field at 4 K for $Co_4Nb_2O_9$. No clear evidence for spin-flop transition was found.

**Fig.3 (a)** Pyroelectric current as a function of temperature under various magnetic fields. **(b)** Calculated electric polarization as a function of temperature at various magnetic fields; The inset shows the polarization versus magnetic field at 8, 22 and 25 K, respectively. **(c)** Electric polarization is plotted as a function of temperature at various poling field. The direction of polarization reversal is also confirmed for negative poling field.

**Fig.4 (a)** Pyroelectric current as a function of time with thermal cyclic in the anti-ferromagnetic state. Such variation can only be observed because of reversible ferroelectric domain condensation. **(b)** Dielectric constant is plotted as a function of temperature. A large peak is recorded across the magnetic transition with increasing magnetic field.

**Table 1**

| Atom | X | Y | Z |
|---|---|---|---|
| Nb1 | 0.00000 | 0.00000 | 0.36000 |
| Co1 | 0.33330 | 0.66670 | 0.02700 |
| Co2 | 0.33330 | 0.66670 | 0.30700 |
| O1 | 0.30500 | 0.00000 | 0.25000 |
| O2 | 0.33330 | 0.29500 | 0.08300 |

**Figure 1**

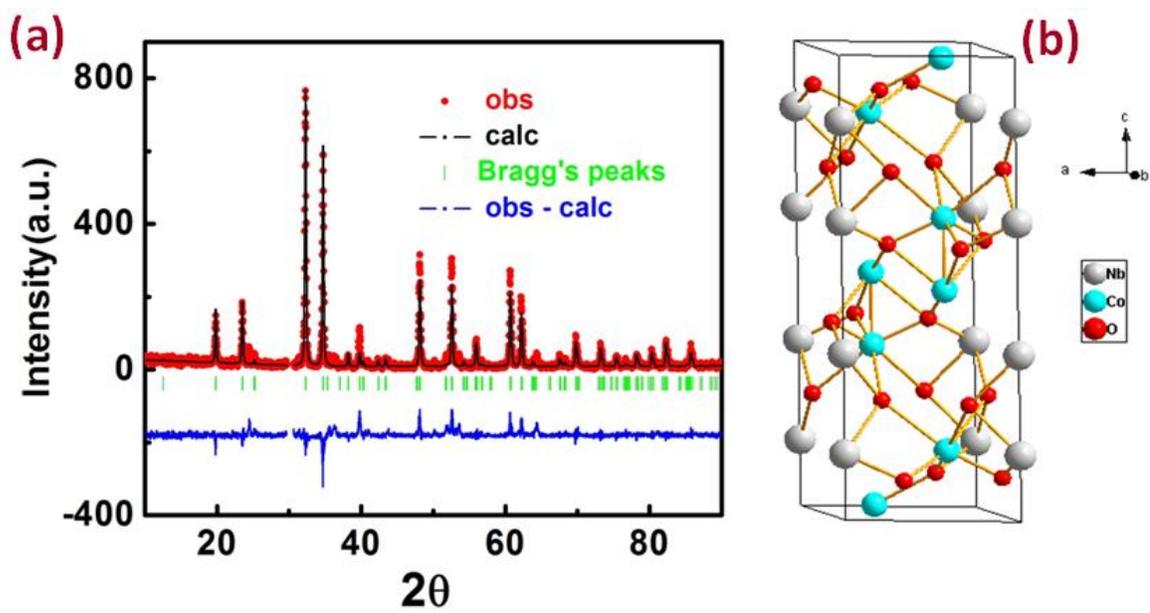

**Figure 2**

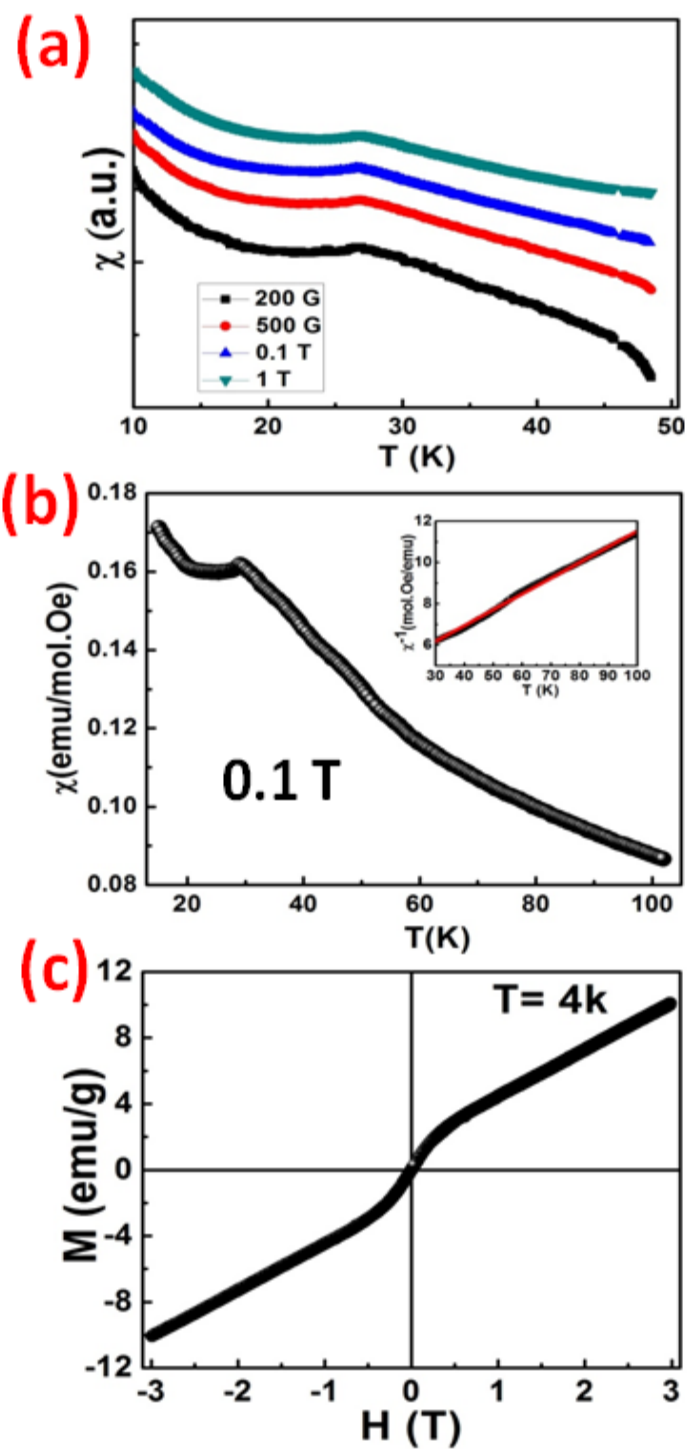

**Figure 3**

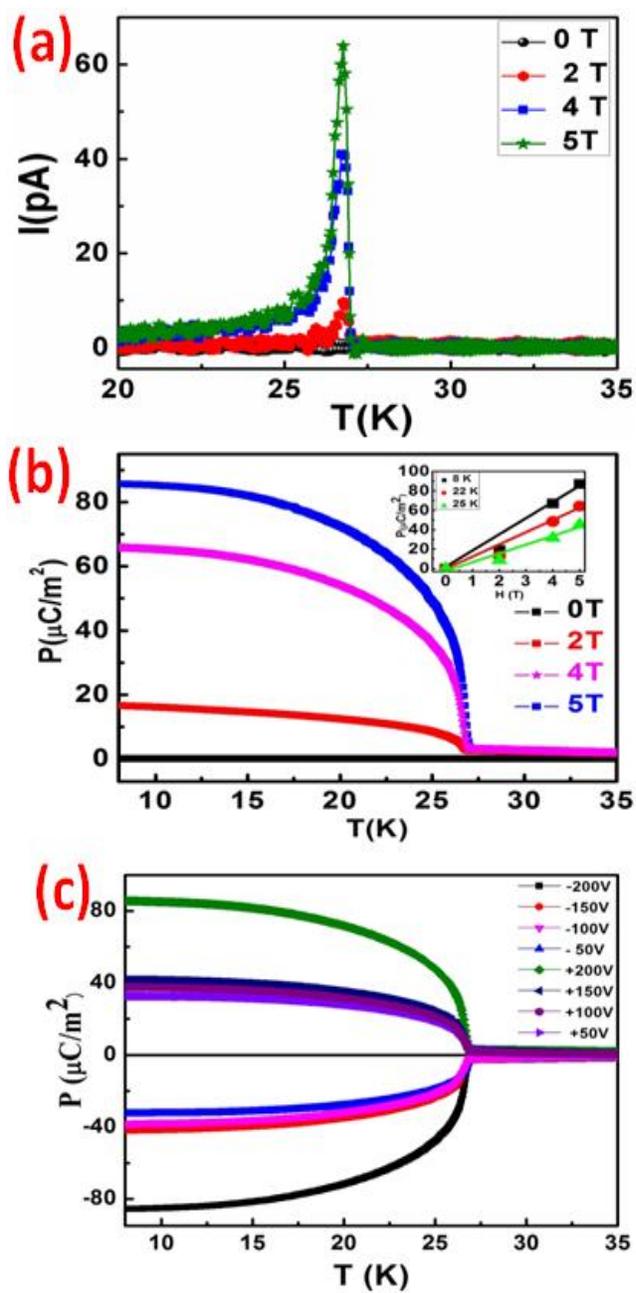

**Figure 4**

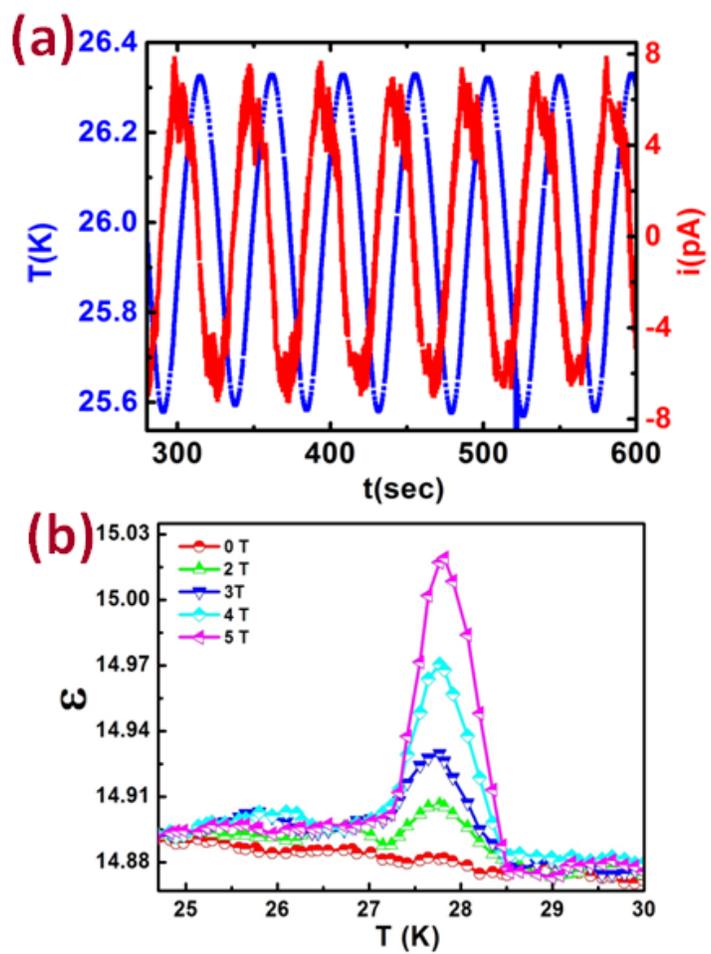